\documentclass{appolb}
\usepackage{psfig}

\begin{document}
\title{Wobbling motion in the multi-bands crossing region:\\
\it dynamical coupling mode between high- and low-$K$ states %
\thanks{Oral presentation at International Conference, {\it High Spin Physics
2001}, Warsaw University, Poland.}%
}
\author{M. Oi$^{a,b}$, A. Ansari$^{c}$, T.Horibata$^{d}$, 
N. Onishi$^{e}$, and P. M. Walker$^{a}$
\address{$^{a}$Department of Physics, University of Surrey, 
Guildford GU2 7XH, U.K.; 
$^{b}$Department of Applied Physics, Fukui University, Fukui
910-8507, Japan; $^{c}$Institute of Physics, Bhubaneswar 751 005,
India; $^{d}$Department of Engineering, Aomori University, Aomori
030-0943, Japan; 
$^{e}$Department of Engineering, Yamanashi University, Kohu
400-8510,Japan}
}
\maketitle
\begin{abstract}
We analyze a mechanism of coupling of high- and low-$K$ bands in terms of a
dynamical treatment for nuclear rotations, i.e.,wobbling
motion. The wobbling states are produced through the generator 
coordinate method after angular momentum projection ({\it
GCM-after-AMP}), in which the intrinsic states are constructed through
fully self-consistent calculations by 
the 2d-cranked (or tilted-axis-cranked) HFB method .
In particular, the phenomena of ``signature inversion'' and ``signature
splitting'' in the t-band (tilted rotational band) are
explained in terms of the wobbling model. 
Our calculations will be compared with new data for in-band
E2 transition rates in $^{182}$Os,
which may shed light on the mechanism of the anomalous
$K=25$ isomer decay, directly to the yrast band.
\end{abstract}
  
\section{Introduction}
 In the $A\simeq 180$ region of the rare-earth nuclei,
three bands are observed to interact with each other at $I\simeq
14\hbar$. The bands are g-, s- and t-bands. 
The g- and s-bands are well-known rotational bands of the ground state
and rotation-aligned configurations, respectively. On the other hand,
the t-band means ``tilted-rotating'' band which was proposed by
Frauendorf \cite{Fr81}.
In $^{184}$Os, the t-band has a $K^{\pi}=10^+$ band head, while
 $^{182}$Os has $K^{\pi}=8^+$. These high-$K$ states are assigned
to two quasi-particle excitations in the neutron i$_{13/2}$ orbits. 
Because the neutron Fermi levels in these nuclei are in the
high-$\Omega$ orbits in the i$_{13/2}$, it is expected that high-$K$
states are excited favourably and appear near the yrast line.

\begin{figure}[bt]
\begin{center}
\psfig{figure=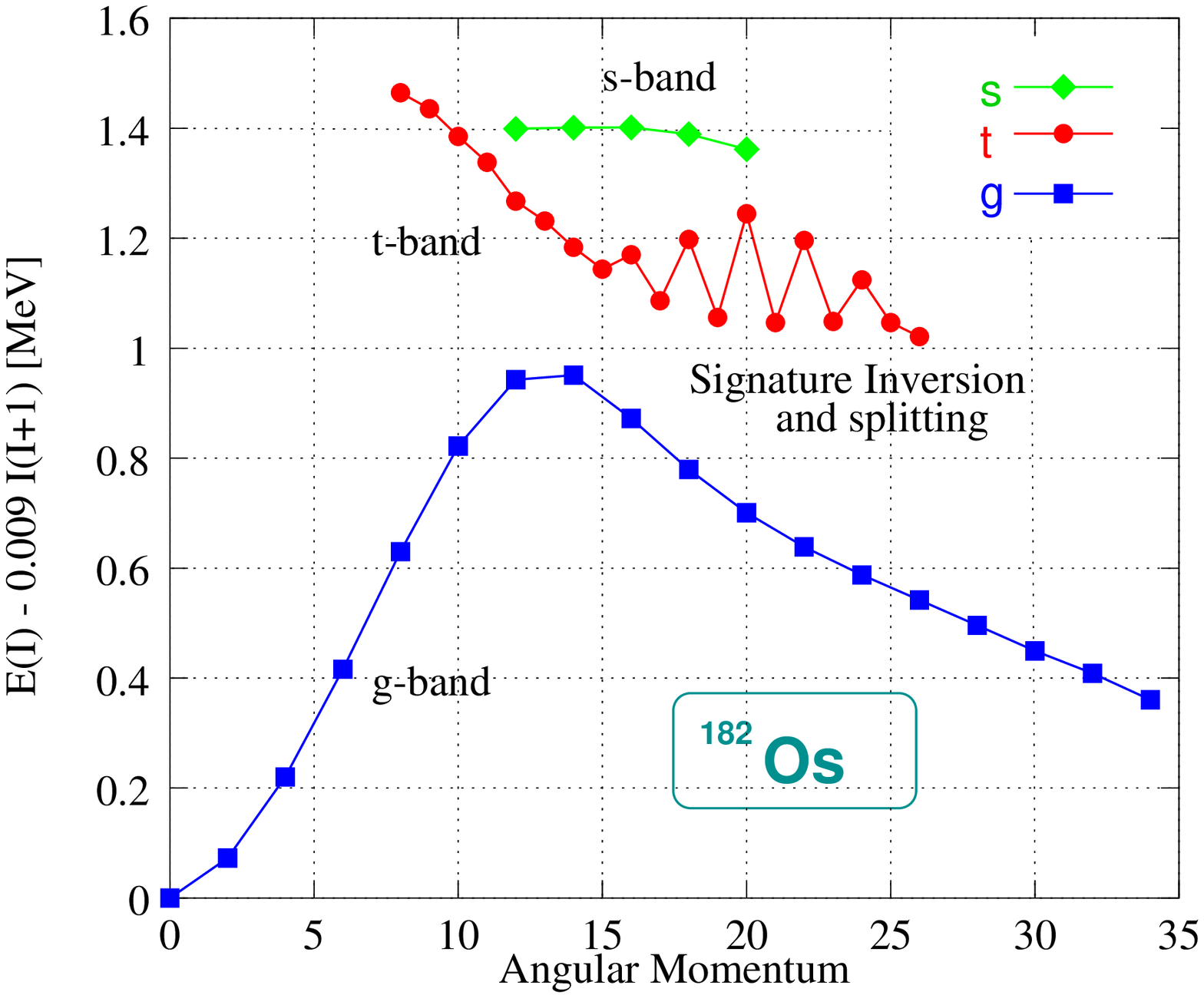,width=12cm,height=5.5cm}
\end{center}
\caption{Multi-bands crossing in $^{182}$Os.}
\label{os182}
\end{figure}

These three bands, i.e., g- and s-bands (low-$K$) and t-band
(high-$K$), seem to interact with each other near $I=14\hbar$
to cause a backbend in the yrast line.
An interesting phenomenon here is that the t-bands split into two
sequences: one of them consists of even-$I$ members while the
other of odd-$I$, which is energetically lower than the former.
(See Fig.\ref{os182}.)
Therefore, it seems that 
so-called ``signature splitting'' and ``signature inversion''
are induced by these multi-band crossings.
These phenomena are observed systematically for even-even nuclei 
in the $A\simeq 180$ region such as $^{180-184}$W and $^{182-186}$Os.
\section{Wobbling model}
	To explain these phenomena, i.e., a coupling between
high- and low-$K$ states ( in other words, break-down of the
$K$-selection rule), we propose the {\it wobbling model}.
The term ``wobbling motion'' was originally introduced into
nuclear structure physics by Bohr and Mottelson \cite{BM75}, 
in order to explain excited rotational structures of 
triaxial nuclei. The wobbling motion in this context is
considered as the small deviation or fluctuation of angular momentum
from the typical collective rotation around the $x$-axis.
The motion is quantized and presents phonon-like structures above the
ground state band.
\begin{figure}[h]
\psfig{figure=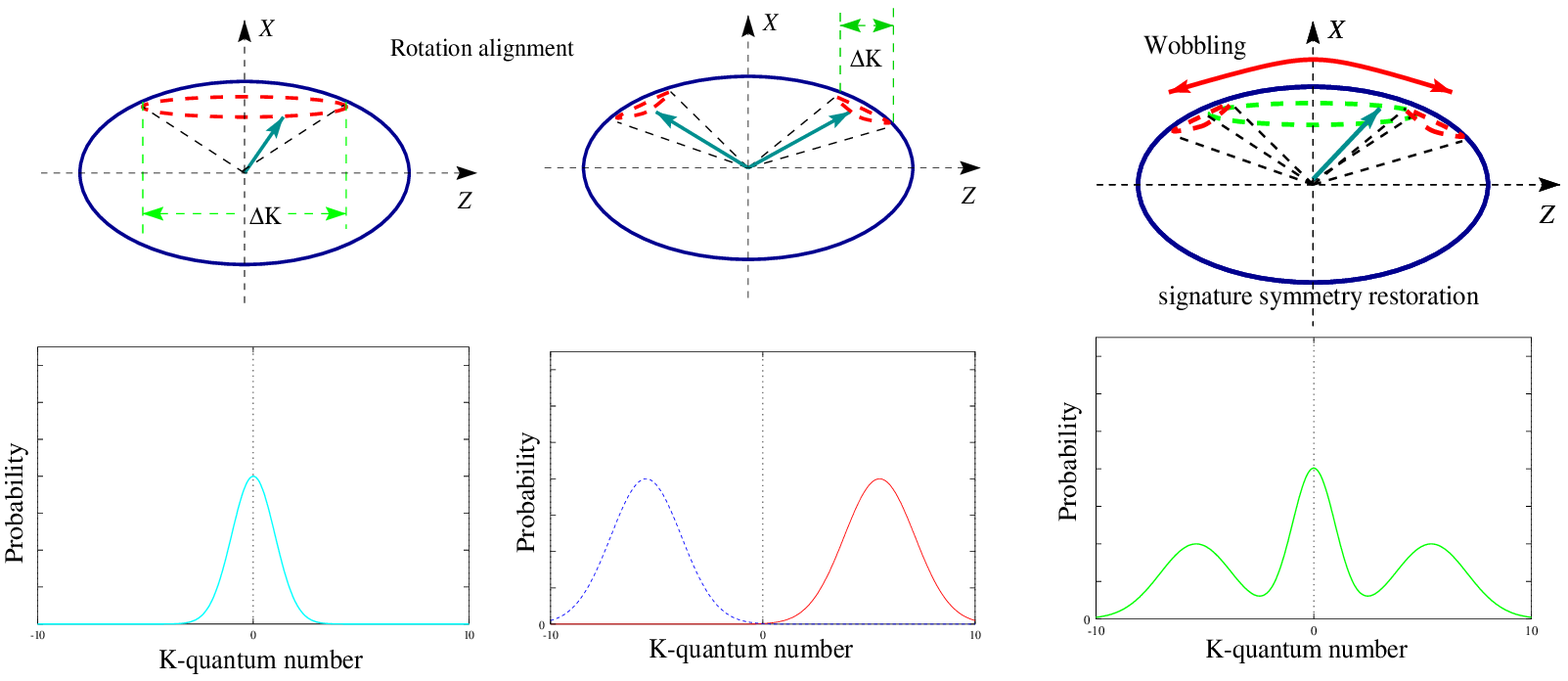,width=11cm,height=4cm}
\caption{Schematic explanation of the wobbling motion}
\label{wobble}
\end{figure}

On the other hand, our wobbling motion is based on the theory by 
Kerman and Onishi \cite{KO81}.
 It is based on the time-dependent variational model
(TDVM) and attempts to handle general three-dimensional nuclear rotations.
In our present model, we describe the wobbling state, $|{\rm
Wbll}\rangle$, as a superposition of 
different tilted-axis-cranked HFB states,
$|{\rm HFB}(\theta, J)\rangle$, where $\theta$ and $J$ are 
defined in the constraints of angular momentum.
$J_x = \langle \hat{I}_x \rangle = J\cos\theta;
J_y = \langle \hat{I}_y \rangle = 0;
J_z = \langle \hat{I}_z \rangle = J\sin\theta$.
(Note that the tilt angle $\theta$ is measured 
from the $x$-axis to see the deviation of the rotation axis 
from the $x$-axis. )

The wobbling states are written,
\begin{equation}
|{\rm Wbbl }^I_M\rangle = {\sum_K \int d\theta} f^I_K(\theta)
\hat{P}^I_{MK}|{\rm HFB}(\theta, J)\rangle,
\end{equation}
where $f^I_K(\theta)$ is the GCM wave function and the variation
is made with respect to this function.
The pairing-plus-Q$\cdot$Q force is employed as an 
residual interaction for the spherical parts (from the Nilsson model)
in the Hamiltonian.
Quantization is made through the angular momentum projection technique 
($\hat{P}^I_{MK}$).

As a result, our wobbling motion can describe not only small 
but also large amplitude wobbling motion around any types of
rotations (with any kinds of deformations). 
In other words, this model describes a dynamical picture
of nuclear excited modes such as rotations and vibrations.

Fig.\ref{wobble} shows schematically how the wobbling motion occurs.
As a nucleus rotates rapidly, the rotation-alignment occurs, in both of
the g- and t-bands. As a result, the yrast band takes on the s-band
character (left panel) and the t-band has some amount of fluctuations in
$K$-distribution as well (center panel). As the nucleus rotates more
rapidly, the $K$ fluctuation around high- and low-$K$ components 
spreads and finally there is an overlap between them (right panel).
As a result, the corresponding state has very large fluctuation in
$K$. Semi-classically, this state coupling low-$K$ states with
high-$K$ can be pictured as the wobbling motion.

\begin{figure}[hbt]
\psfig{figure=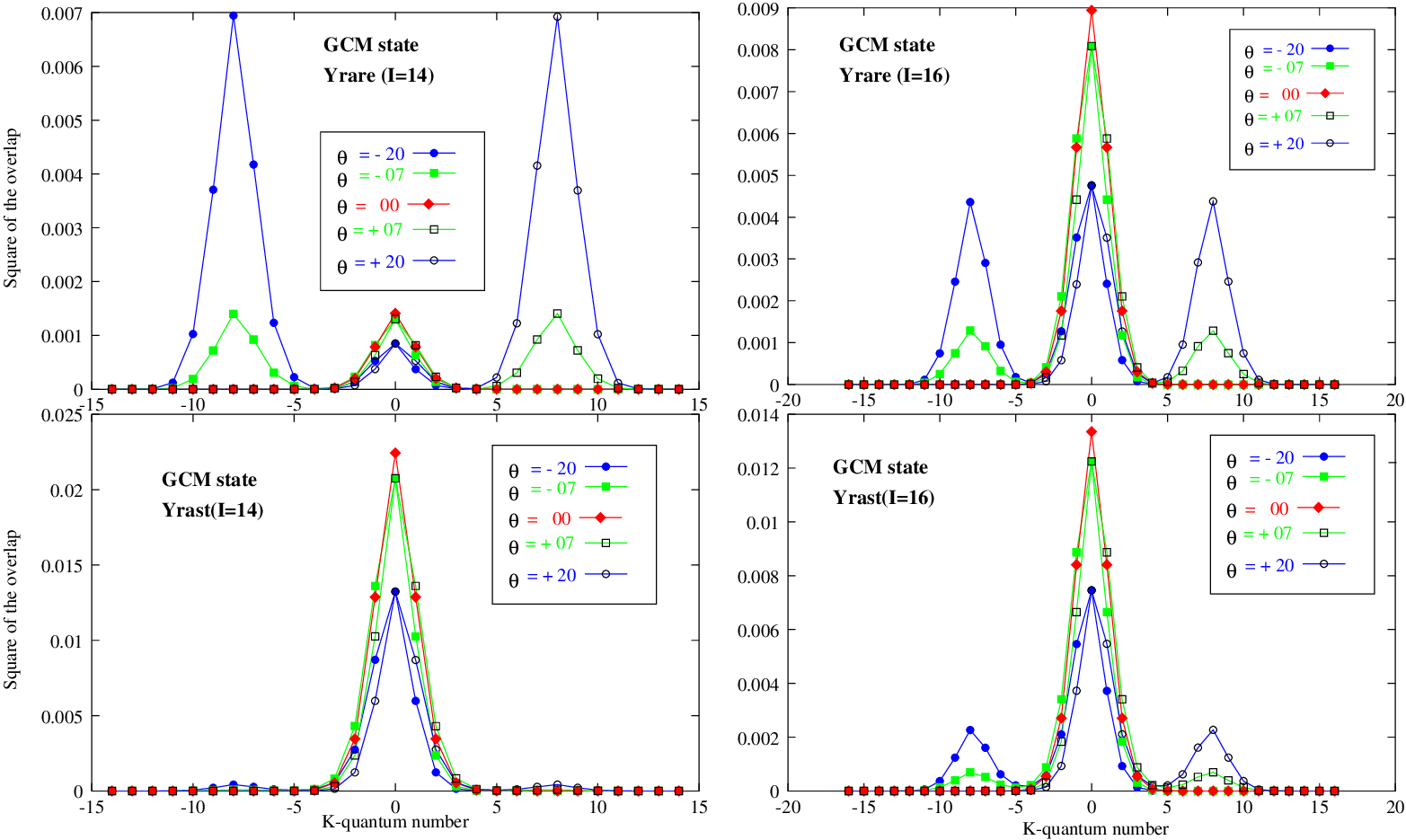,width=12cm,height=5.5cm}
\caption{Overlap $\langle{\rm HFB}(\theta,J)|\hat{P}^{I \dag}_{KK}|{\rm
Wbbl}^I_M\rangle$ for $I=14\hbar$ (before crossing) and $I=16\hbar$
(after crossing). See Ref.\cite{OAHO00} for details.}
\label{calc}
\end{figure}

We performed numerical calculations and confirmed that the wobbling model
can describe the main features of the multi-bands crossing phenomena.
Particularly, our calculations can reproduce the coupling of low- and
high-$K$ components in the yrast and excited bands through the crossing.
(See Fig.\ref{calc}.)
Then, we propose an interpretation 
for the mechanism of the signature inversion through
the wobbling model: first, as a result of the rotation-alignment, the
yrast line takes on the s-band character; then, the t-band approaches 
to the yrast line and inter-band interactions start 
between even-$I$ members of the t-band and members in the s-band,
while odd-$I$ members in the t-band are unperturbed because there are
no counterparts in the yrast band. Thus, the even-$I$ members are
pushed up from the original t-band sequence 
while the odd-$I$s keep staying in the sequence.
This mechanism can explain the signature inversion (and splitting) 
phenomena.
At the same time, the yrast line is pushed down through the 
inter-band interaction, and as a consequence backbending 
can be enhanced.
The detailed results and discussions are presented in our previous paper
\cite{OAHO00}.
\section{Experimental investigations of the wobbling motion}
In $^{182}$Os, there is an isomer with $K^{\pi}=8^-$, and
its rotational band seems to be isolated.
Thus, it is expected to have a pure high-$K$ ($K=8\hbar$) structure 
(at least, no low-$K$ component). Podoly\'ak et al. \cite{Zs01} 
have measured the effect of low-$K$ components in the t-band ($K^{\pi}=8^+$)
by the comparison of the transition rates between these two
high-$K$ bands (i.e.,$K^{\pi}=8^+$ and $8^-$).

\begin{figure}[hbt]
\psfig{figure=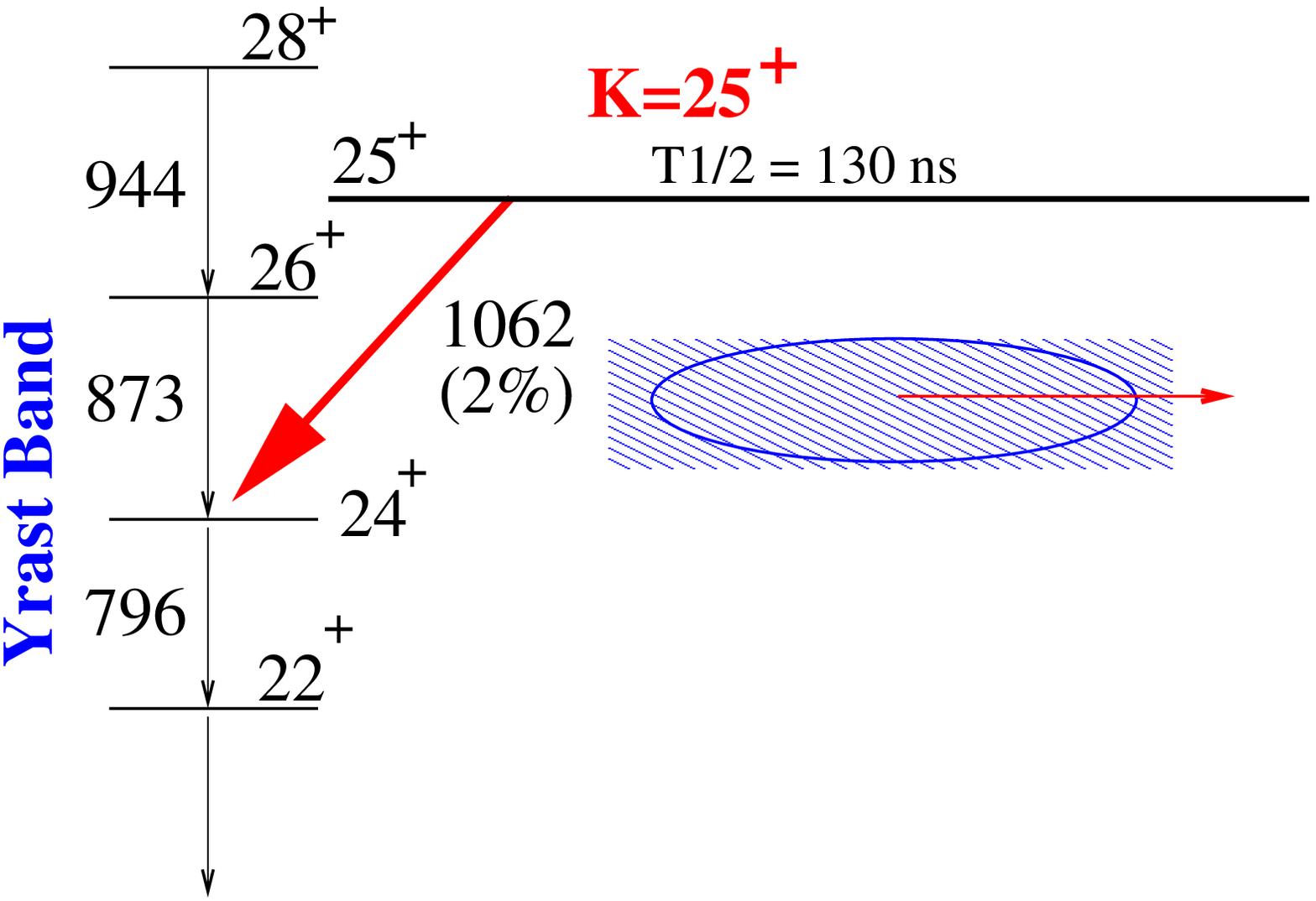,width=12cm,height=4.5cm}
\caption{Transition from $K^{\pi}=25^+$ to the yrast line \cite{Ch88}.}
\label{exp1}
\end{figure}

Furthermore, it is hoped to identify the
effect of high-$K$ components ($K\simeq 8\hbar$) in the
yrast line at high spin ($I\simeq 24\hbar$),
which influences the transition probability from the $K^{\pi}=25^+$
isomer (having a very short half life, $T_{1/2}=$130 ns) to the yrast line.
(See Fig.\ref{exp1}.)

\section{Summary}
We have investigated a mechanism of  signature
inversion and splitting in the $A\simeq 180$ region by 
means of {\it GCM-after-AMP} on the tilted-axis-cranked HFB states.
With this method, we have qualitatively reproduced the main features of
the excited level structure in the high-$K$ t-band.
We interpret this result from a point of view of an inter-band
interaction between the low-$K$ (s-) and high-$K$ (t-) bands.
We have shown that the perturbed states have the character of 
wobbling motion, that is, a dynamical mode coupling low-$K$ PAR
(principal-axis
rotation) and high-$K$ TAR (tilted-axis-rotation) states.
In terms of this wobbling model, we have discussed an enhancement of
 backbending in the $A\simeq 180$ region, in which the typical
rotation-alignment is somewhat suppressed due to the location of the
neutron Fermi level.

\end{document}